\documentclass[aps,prl,twocolumn,amsmath,amssymb,floatfix,
superscriptaddress]{revtex4}
\usepackage{subfigure}
\usepackage{graphicx}

\def\avg#1{\left\langle#1\right\rangle}

\def\ket#1{\left|#1\right\rangle}

\def\be{\begin{equation}}       \def\ee{\end{equation}}
\def\bea{\begin{eqnarray}}      \def\eea{\end{eqnarray}}
\def\ba{\begin{array} }
\def\ea{\end{array} }
\def\bnum{\begin{enumerate} }
\def\enum{\end{enumerate}}

\def\nn{\nonumber}

\def\=>{\Rightarrow}
\def\>{\rightarrow}
\def\A{\uparrow}
\def\V{\downarrow}

\def\eye2{Fathbb{I}}

\def\Eq#1{Eq.~(\ref{#1})}
\def\Fig#1{Fig.~\ref{#1}}

\def\eff{\mathrm{eff}}

\begin{document}
\title{
Fragile Mott Insulators}
\author{Hong Yao}
\affiliation{Department of Physics, University of California,
Berkeley, CA 94720, USA} 
\affiliation{Materials Sciences Division,
Lawrence Berkeley National Laboratory, Berkeley, CA 94720, USA}
\author{Steven A. Kivelson}
\affiliation{Department of Physics, Stanford University, Stanford,
CA 94305, USA}
\date{\today}
\begin{abstract}
We prove that there exists a class of crystalline insulators, which we call ``fragile Mott insulators''
which are not adiabatically connected to any sort of band insulator provided time-reversal and certain
point-group symmetries are respected, but which are otherwise unspectacular in that they exhibit no
topological order nor any form of fractionalized quasiparticles. Different fragile Mott insulators are
characterized by different nontrivial one-dimensional representations of the crystal point group. We
illustrate this new type of insulators with two examples: the $d$-Mott insulator discovered in the checkerboard
Hubbard model at half-filling and the Affleck-Kennedy-Lieb-Tasaki insulator on the square lattice.
\end{abstract}
\maketitle

Crystalline insulators are called ``band insulators'' if they can be adiabatically deformed into a system of non-interacting electrons in which all the occupied electronic bands are separated from empty bands by a finite energy gap. Insulating phases with spontaneously broken symmetries are clearly distinct phases of matter, which cannot evolve into a band insulator without undergoing a phase transition. In addition, however, there  
can exist insulating phases of crystalline systems with no broken symmetries which non-the-less cannot be adiabatically related to a band insulator - such phases are often referred to as ``Mott insulators'' \cite{mott1968}. 
 
Exotic ``featureless Mott insulators''  or ``quantum spin liquids'' were first proposed \cite{anderson1973} as candidate ground states for highly frustrated quantum antiferromagnetic insulators. The existence of fractionalized excitations \cite{anderson1987,kalmeyer1987} and/or topological order (if 
all the excitations are fully gapped) \cite{kivelson1987,wen1989b} are the defining features of such phases. For instance, in the 2D RVB state \cite{anderson1973, anderson1987, kalmeyer1987, kivelson1987}, the spin and charge 
of an electron are separated in the low energy excitation spectrum. The existence of such 
 phases in various 
exactly solvable (even though often contrived) models has now been established (see e.g. Ref. \cite{moessner2001, kitaev2006
}), but while there is increasingly interesting evidence of their existence in specific materials \cite{lee2008,balents2010}, it is fair to say that there is still no single material that has been clearly {\em proven} to possess a spin liquid ground state.

In this paper, we define a fourth class of crystalline insulators, which are qualitatively different from the three discussed above, which we propose to call ``
fragile Mott insulators.'' 
The ground state of such an insulator transforms non-trivially under the operations of the point group.  From this it follows that, so long as the allowed adiabatic paths respect time reversal symmetry and the crystalline point group symmetries, 
fragile Mott insulators cannot be adiabatically connected to any band insulator; there must be at least one phase transition along any path connecting a 
fragile Mott insulator to a band insulator. (For some but not all point groups, it is not even necessary to assume time reversal symmetry to distinguish  fragile Mott insulators [11].)  On the other hand,  
fragile Mott insulators are qualitatively different from quantum spin liquids in the sense that there are no fractionalized excitations and they have a unique ground state on the torus. Here we use the term ``
fragile'' in describing these phases because they can be sharply  
distinguished from band insulators only when certain crystalline point group symmetries are preserved. 
We give two examples of solvable models of interacting electrons which have 
fragile Mott insulating phases.  
The protection of topological order by crystalline symmetries was previously studied (see e.g. Ref. \cite{wen2002, berg2008, gu2009, pollmann20092010,turner2009}). For instance, 
its transformation properties under spatial inversion 
distinguishes  the Haldane phase of a spin-1 chain from a trivial phase \cite{berg2008, gu2009, pollmann20092010}.   

To establish that 
fragile Mott insulators are a distinct phase of matter, we rely on the following theorem:

{\bf Theorem:} {\it In any time reversal invariant band insulator, the ground state must transform according to the identity representation of the crystal point group. }

{\bf Proof:} Consider a band insulator described by a non-interacting Hamiltonian ${\cal H}$ where the fermionic operator $\phi^\dag_n$ creates a single particle energy eigenstate $\ket{\phi_n}\equiv \phi^\dag_n\ket 0$. 
Time reversal symmetry is represented by an anti-unitary operator $\Theta$, where $[\Theta,{\cal H}]=0$ and $\Theta^2=-1$, so all single-particle 
eigenstates form Kramers doublets, {\it i.e.}  
$\Theta \phi_n \Theta^{-1} = \phi_{\bar n}$ and $\Theta \phi_{\bar n} \Theta^{-1} = -\phi_{n}$,  
where $\ket{\phi_n}$ and $\ket{\phi_{\bar n}}$ 
are linearly independent states.  Since Kramers doublets are either both occupied or both unoccupied in a band insulator, any band insulator can be described by the wave function 
\bea 
\ket{\Psi_0}=\prod_{(n,\bar n)\in \textrm{occ.}} \left[\phi^\dag_n \phi^\dag_{\bar n}\right]\ket 0,
\eea
where ``$\textrm{occ.}$'' signifies the set of occupied states. Let $C\in G$ be an element of the point group, $G$. Since $[C,{\cal H}]=0$,  any single particle eigenstate $\ket{\phi_n}$ can be chosen to be a simultaneous eigenstate of $C$, {\it i.e.} $C\ket{\phi_n}=\lambda_n\ket{\phi_n}$. 
Suppose that the degree of $C$ is $m\geq 1$, namely $C^m=1$, which indicates that its possible eigenvalues are $\lambda_n=e^{i2\pi j_n/m}, j_n=1,\cdots,m$. 
Since spatial symmetries commute with 
 time reversal symmetry, $[C,\Theta]=0$, 
it follows that $C\ket{\phi_{\bar n}}=C\Theta\ket{\phi_n}=\Theta C\ket{\phi_n}=\lambda_n^\ast \Theta\ket{\phi_n}=\lambda^\ast_n \ket{\phi_{\bar n}}$,  
so $\lambda_{\bar n}=\lambda^\ast_n$. With this, we obtain
\bea 
C\ket{\Psi_0}=\prod_{(n,\bar n)\in \textrm{occ.}}\left[\lambda_n \lambda_{\bar n}\right]\ket{\Psi_0}=\ket{\Psi_0}.
\eea    
Q.E.D.
Note that the above proof is valid for any finite lattice size, and remains true as the thermodynamic limit  
is approached.

\begin{figure}[b]
\subfigure[]{
\includegraphics[scale=0.22]{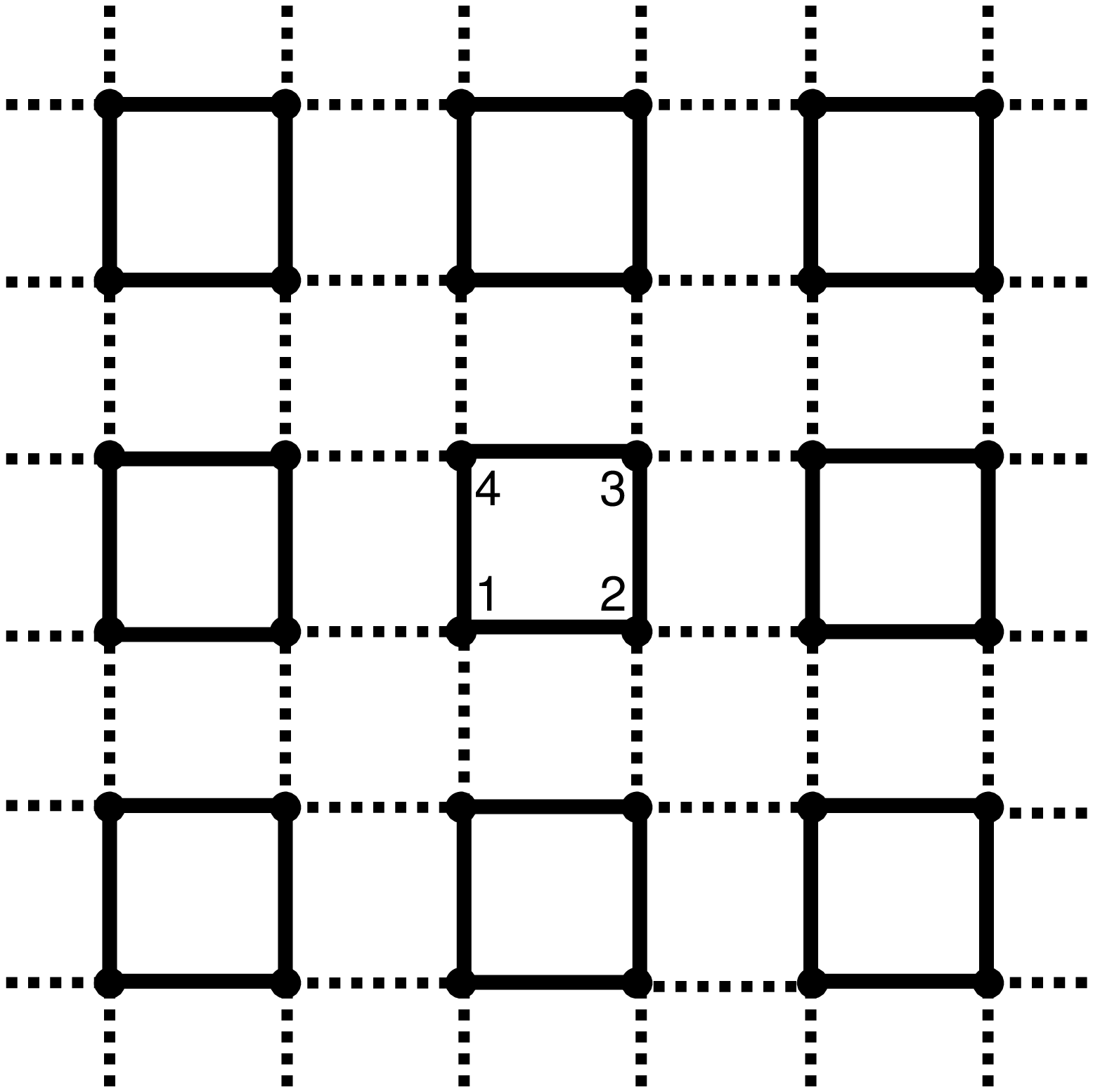}}~~
\subfigure[]{\includegraphics[scale=0.31]{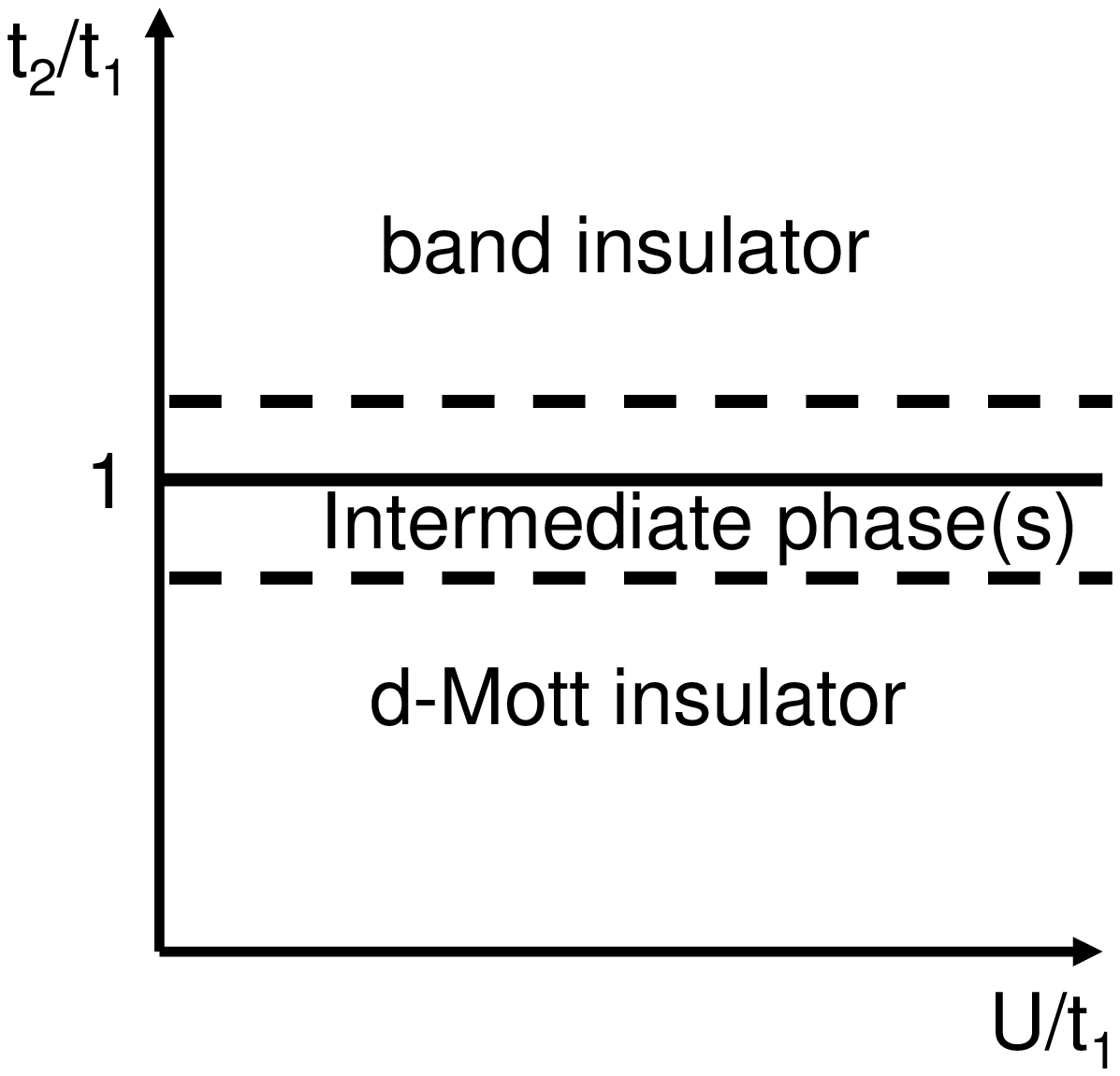}}
\caption{\!(a) Schematic representation of the checkerboard lattice. Intraplaquette hoppings are $t_1$ between nearest neighbors and $t_2$ along diagonal direction and interplaquette $t'$. (b) Phase diagram of the Hubbard square (solid lines) and of the checkerboard Hubbard model (dashed lines) in the small $t^\prime$ limit.}
\label{fig:checkerboard}
\end{figure}

{\it The Hubbard Square:}  The simplest illustration of the physics of this new phase comes  when we consider the ground-state of the four-site Hubbard model with the $C_{4v}$ symmetries of a square, for which the most general Hamiltonian in the absence of spin-orbit coupling is
\bea
\label{Hsq}
H^{sq}&=& \sum_{j=1}^4\sum_\sigma \left\{ -t_1 \left[c_{j,\sigma}^\dagger c_{j+1,\sigma} + {\rm h.c.}\right] \right . \\
&& \left .+t_2 c_{j,\sigma}^\dagger c_{j+2,\sigma} +Uc_{j,\uparrow}^\dagger c_{j,\downarrow}^\dagger c_{j,\downarrow} c_{j,\uparrow}\right\}
\nonumber
\eea
where $c_{j,\sigma}^\dagger \equiv c_{j+4,\sigma}^\dagger$ creates an electron with spin-polarization $\sigma$ on site $j$. The ground-state phase diagram \cite{scalapino1996,schumann2002} of this ``Hubbard molecule'' with 4 electrons (in average one electron per site) is shown in \Fig{fig:checkerboard}(b).  

For $t_2>t_1$, the ground-state is unique, has spin 0, and transforms trivially under the operations of the symmetry group, {\it i.e.} it has s-wave symmetry.  There is a gap to the first excited state.  At $U = 0$,  the ground-state of the non-interacting model is non-degenerate so, in the sense of adiabatic continuity, this  can be classified as a band insulator, and it is so labeled.  

For $t_1 > t_2$ and $U>0$, the ground-state is unique, has spin 0, and there is a gap to the first excited state. However, the ground-state is odd under rotation by $\pi/2$ and under reflection through various mirror planes - it has the symmetry of the $d_{x^2-y^2}$ representation of the point-group.  For $U=0$, the ground-state is 6-fold degenerate, but the structure of the ground-state in the $U/t_1\to 0$ limit, as we will discuss below, is well defined, and is readily seen to be an entangled state that cannot be expressed as a single Slater determinant.  Because of its d-wave symmetry, we call this phase a ``d-Mott'' state.  

The phase boundary between these two phases occurs at $t_1=t_2$.  At this point, the Hubbard square has a higher symmetry - the symmetric group of degree four $S_4$ or the symmetry of a tetrahedron. Consequently, the d-wave and s-wave ground states  combine to form 
a two dimensional representation 
of the symmetric group $S_4$; 
the two states are degenerate by symmetry.

While the Hilbert space in this problem is small enough that it is easily solved, 
it is useful to understand the nature of the phases approximately in various limits.  In the large $U/t_1$ limit, the problem is equivalent to the spin-1/2 Heisenberg square, with exchange coupling $J_1=4t_1^2/U$ and $J_2=4t_2^2/U$ between nearest and next-nearest neighbor spins, respectively.  There are two spin-0 states, 
\bea
&&|\Psi_s\rangle = b_{13}^\dagger b_{24}^\dagger\  |0\rangle  
\\
&&|\Psi_d\rangle = \frac{1}{\sqrt3} \left[ b_{12}^\dagger b_{34}^\dagger - b_{14}^\dagger b_{23}^\dagger \right]\ | 0\rangle
\nn
\eea
where $b_{ij}^\dagger \equiv [c_{i,\uparrow}^\dagger c_{j,\downarrow}^\dagger + c_{j,\uparrow}^\dagger c_{i,\downarrow}^\dagger]/\sqrt{2}$ creates a singlet pair between sites $i$ and $j$.  Each of these states is an eigenstate of the Hamiltonian, since $|\Psi_s \rangle$ is the unique spin-0 state which transforms according to the identity representation of the point group, while $|\Psi_d\rangle$ is the unique state with $d_{x^2-y^2}$ symmetry.    
While $|\Psi_s\rangle$ is a simple valence bond state, $|\Psi_d\rangle$ is a ``short-ranged resonating valence bond state,'' along the lines that were considered in 
Ref. \cite{anderson1973,kivelson1987}. 

Still more interesting is the small $U$ limit.  For $U=0$ and $t_2 > t_1$, the ground-state is readily seen to be the simple Slater determinant state,
\bea
|\Psi_s\rangle = a_{\pi/2,\uparrow}^\dagger a_{\pi/2,\downarrow}^\dagger a_{-\pi/2,\uparrow}^\dagger a_{-\pi/2,\downarrow}^\dagger |0\rangle
\eea
where $a_{k,\sigma}=(1/2)\sum_j e^{ik j}c_{j,\sigma}^\dagger$.
However, for $t_1 > t_2$, the ground state in the $U/t_1\to 0$ limit is the entangled state
\bea
|\Psi_d\rangle \to \frac 1 {\sqrt{2}} a_{0,\uparrow}^\dagger a_{0,\downarrow}^\dagger \left[a_{\pi/2,\uparrow}^\dagger a_{\pi/2,\downarrow}^\dagger - a_{-\pi/2,\uparrow}^\dagger a_{-\pi/2,\downarrow}^\dagger \right] |0\rangle .
\eea
The fact that even in the $U/t_1 \to 0$ limit, the ground-state is entangled is a remarkable feature of this particular Hubbard molecule \cite{chakravarty2001}, which is directly related to the existence of the d-Mott phase.

{\it The d-Mott insulator}: The existence of a d-Mott phase in an extended system follows simply from the solution of the Hubbard square \cite{yao2007b}. 
Consider the ``checkerboard Hubbard model,'' shown in \Fig{fig:checkerboard}(a), which we can think of as a model of a molecular crystal built of Hubbard squares \cite{tsai2006,yao2007b,kocharian2006}.  Now, $c^\dagger_{a,j,\sigma}$ creates an electron on the $j^{th}$ site of the $a^{th}$ molecule with spin polarization $\sigma$, and
\bea 
H=\sum_a H^{sq}_a -t^\prime \sum_{\langle aj,a^\prime j^\prime\rangle}\sum_\sigma\left [ c^\dagger_{a,j,\sigma} c_{a^\prime,j^\prime,\sigma} + {\rm h.c.}\right],
\eea  
where $H^{sq}_a$ is the molecular Hamiltonian defined as in \Eq{Hsq} with $c^\dagger_{j,\sigma}\to c^\dagger_{a,j,\sigma}$ and $\langle aj,a^\prime j^\prime\rangle $ signifies pairs of nearest-neighbor sites on adjacent molecules.  
In the  $t'\to 0$ limit, the system consists of disconnected square molecules so the ground-state is given by the tensor product of the ground-states of the isolated squares.   
It follows from the existence of a gap in the spectrum 
that the thermodynamic properties of the system evolve smoothly  for non-zero $t^\prime$ as long as $t'$ is small enough and can be computed perturbatively.  
In particular, all symmetry properties of the ground state are independent of $t^\prime$ in this regime, so with the possible exception of a region of width $\sim t^\prime/t_1$ about the line $t_1=t_2$ (which we discuss separately, below), the phase diagram in \Fig{fig:checkerboard}(b) applies to the thermodynamic limit of the checkerboard Hubbard model for small enough 
but non-zero $t^\prime/t_1$.

It remains to establish, for the extended system, whether or not each phase can be adiabatically connected to a non-interacting insulator.
The band-structure of this model with $U=0$ is easily computed.  For $t_2 - t_1 > |t^\prime|$, there is a band-gap  separating the occupied and empty states.  As this state is approached smoothly as $U\to 0$, it is still clearly the case that the label ``band-insulator'' applies to the large $t_2/t_1$ phase.  

To prove that the d-Mott phase at $t_1 - t_2 > |t^\prime|$ is not a band-insulator, we consider a system of $L$ by $L$  
molecules with periodic boundary conditions and $L$ odd. Then, it is obvious that the ground state wave function acquires a minus sign under $90^\circ$ rotation ($C_4$) around a plaquette center or reflections (labeled by $\sigma_d$
) along a diagonal line across  
the plaquette centers, 
\bea 
\sigma_{d
}\ket{\Psi_{d\textrm{-Mott}}} =-\ket{\Psi_{d\textrm{-Mott}}},
C_4\ket{\Psi_{d\textrm{-Mott}}} =-\ket{\Psi_{d\textrm{-Mott}}},
\eea
which implies the ground state wave function transforms like a $d_{x^2-y^2}$ orbital under the point group $C_{4v}$. 
According to our theorem, this implies that the d-Mott phase cannot be adiabatically related to a band insulator.  If $L$ is even, the ground-state transforms according to the trivial representation of the point group.  Thus, for these boundary conditions, we cannot use symmetry to prove that the system is not adiabatically related to a band insulator.  However, the identity of a phase should not depend on the way the thermodynamic limit is approached.  Thus, we believe that proving that adiabatic evolution to a band insulating state is impossible for $L$ odd is sufficient for establishing that this is a distinct phase \cite{footnote1}. 

To gain further insight into this problem, we focus on the behavior of the system near the boundary between the band and d-Mott insulating phases.  Here, for $t^\prime=0$ and $t_1=t_2$, there are two low-lying states per molecule, so to study the effects of non-zero coupling $t'$ between molecules and non-zero $|t_1-t_2|$, we must derive an effective Hamiltonian using near-degenerate perturbation theory.  We thus define pseudo-spin operators associated with each molecule, where $\tau_a^z = 1$ if molecule $a$ is in its s-wave ground-state, and  $\tau_a^z = -1$ if it is in its d-wave ground state.  We define the raising operator $\tau_a^+$  
that promotes the molecule from its d-wave to its s-wave ground state, etc.  In terms of these operators, we derive   
a pseudo-spin $\frac12$ quantum Ising model defined on the molecular lattice to lowest order in $|t_1-t_2|$ and to second order in $t^\prime$:
\vspace{-0.1cm}
\bea 
H_\eff&=&\sum_{a}J_\eff \Big[ (\vec\tau_a\cdot \hat e_+)(\vec\tau_{a+\hat x}\cdot \hat e_+)\nn\\
&&+(\vec\tau_a\cdot \hat e_-)(\vec\tau_{a+\hat y}\cdot \hat e_-)\Big]
-h\sum_a \tau^z_a,
\label{eq:effH}
\eea
where $\hat e_{\pm}= \frac{\sqrt3}2\hat x\pm\frac12 \hat z$ are unit vectors, $J_\eff=\frac{t'^2}{t_1} g_1(U/t_1)$,  
and $h=(t_2-t_1) g_2(U/t_1)+\frac{t'^2}{t_1}g_3(U/t_1)$   
where $g_{1,2,3}$ 
are regular non-negative functions of $U/t_1$ which can be computed numerically.  
Under a $C_4$ rotation of the system, the lattice points transform as $a\to a^\prime$, and $\vec \tau_a \to e^{i\pi\tau^z/2}\vec 
\tau_{a^\prime} e^{-i\pi\tau^z/2}$;  
the Hamiltonian is invariant under this transformation. 
Any ordered state with a non-zero expectation value of $\tau_a^x$ breaks the global $C_4$ symmetry of the lattice. For $h=0$, the model has two additional global Ising symmetries: $\vec \tau_a \to e^{i\pi\tau^y/2}\vec \tau_{a}e^{-i\pi\tau^y/2}$ and $\vec \tau_a \to e^{i\pi\tau^x/2}\vec \tau_{a^\prime}e^{-i\pi\tau^x/2}$.  
For $h \gg J_\eff$, the ground state is polarized with $\tau_a^z=1$ (band insulator) while for $h \ll - J_\eff$, the ground state has $\tau_a^z=-1$ (d-Mott). We do not, as yet, have definitive results on the nature of the transition(s) between these two extremal phases for small $|h|$.  Based on the obvious similarity between $H_{\eff}$ and the transverse field Ising antiferromagnet, it seems likely that there is at least one intermediate phase with  
spontaneously broken translational and $C_4$ symmetry 
occurs in a region of width $\sim {t^\prime}^2/t_1$ about the $t_1=t_2$ line as shown by the dashed lines in \Fig{fig:checkerboard}(b).

{\it The AKLT insulator}: As a second example, we will consider a model with a low energy effective theory equivalent to the famous AKLT model of quantum antiferromagnetism \cite{affleck1988}.  
For simplicity, we define the model on the same checkerboard lattice as shown in Fig. 1a with a mean electron density of one electron per site (4-electrons per unit cell), but with the Hamiltonian
\bea 
H=K\sum_a
(6-\mathbf{S}_a^2) 
+J\sum_{\avg{a,a^\prime}} P_4(a,a^\prime)+\Delta\sum_a(S^z_a)^2,
\label{eq:AKLT}
\eea
where the total spin on square $a$ is $\mathbf{S}_a=\sum_{j=1}^4 \mathbf{S}_{a,j}$ ($\mathbf{S}_{a,j}=\frac12\sum_{\sigma,\sigma^\prime} c_{a,j,\sigma}^\dagger \mathbf{\tau}_{\sigma,\sigma^\prime}c_{a,j,\sigma^\prime}$)  
and $P_4(a,a') $ is the projection operator onto spin-$(\mathbf{S}_a+\mathbf{S}_{a'})^2=S(S+1)$  
with $S=4$. This Hamiltonian has all the same spatial symmetries as the checkerboard lattice.
 
For $K$, $J > 0$, and $\Delta=0$, this Hamiltonian is positive semi-definite.  The $K$ term is minimized by any state with 4 electrons in a spin 2 state on each molecule. In the ground-state subspace defined by this $K$ term, 
the $J$ term in \Eq{eq:AKLT}, is precisely the spin 2 AKLT Hamiltonian on the square lattice \cite{affleck1988}. The zero energy  ground state is thus seen to be   
\vspace{-0.3cm}
\bea 
&&\ket{\Psi_\textrm{AKLT}}=\frac{1}{{\cal N}}\sum_{\{(j_{a,1},j_{a,2},j_{a,3},j_{a,4})\}}\bigg[\prod_a (-1)^{\delta_a} \nn\\ &&~~~\times\prod_{a} b^\dag_{a, j_{a,1};a+\hat x,j_{a+\hat x,2}}b^\dag_{a, j_{a,3};a+\hat y,j_{a+\hat y,4}}\ket 0\bigg],
\label{eq:wfAKLT}
\eea
where $(j_{a,1},j_{a,2},j_{a,3},j_{a,4})$ labels one of the 24 possible permutations of $(1,2,3,4)$ on each plaquette $a$ and $(-1)^{\delta_a}$ is the signature of permutation. Here, $b^\dag_{a,j_{a,1};a+\hat x,j_{a+\hat x,2}}=\left[c^\dag_{a,j_{a,1},\A} c^\dag_{a+\hat x,j_{a+\hat x,2},\V}+c^\dag_{a+\hat x,j_{a+\hat x,2},\A} c^\dag_{a,j_{a,1},\V}\right]/\sqrt2$ creates a singlet pair between two sites $(a,j_{a,1})$ and $(a+\hat x,j_{a+\hat x,2})$ and $b^\dag_{a,j_{a,3};a+\hat y,j_{a+\hat y,4}}$ is defined similarly. ${\cal N}$ is a normalization constant. Again,  
for the system on a $L$ by $L$ lattice with $L$ odd, we obtain  
\bea 
\sigma_{v(d)}\ket{\Psi_\textrm{AKLT}}&=& - \ket{\Psi_\textrm{AKLT}},
\eea
which indicates that the AKLT insulator on the checkerboard lattice transforms as the 
$A_2$-representation under the $C_{4v}$ group and thus cannot be adiabatically connected to any band insulator. 
Moreover, since the d-Mott insulator transforms in the  
$B_1$-representation ($d_{x^2-y^2}$), as discussed previously,  
the AKLT insulator and d-Mott insulator are distinct states of matter. Different 
fragile Mott insulators are distinguished by the different one dimensional representations of the  point group in question.  

For $K$, $\Delta>0$, and $J=0$, the system becomes disconnected squares, each of which has a unique ground state with $S=2$ and $S^z=0$, and the system's ground state is then given by the direct product state $\ket{{\cal D}}=\prod_a \otimes\ket{S_a=2,S^z_a=0}$. Due to the existence of a gap in the spectrum, the qualitative nature of $\ket{{\cal D}}$ survives for non-zero but small enough $J$.  Since $\ket{{\cal D}}$ on a $L$ by $L$ lattice with $L$ odd transforms like the $d_{x^2-y^2}$-representation under $C_{4v}$, it is adiabatically connected to the d-Mott state but not to the AKLT insulator.

{\it Concluding remarks}:  Beyond the point of principle - that 
fragile  
Mott insulators are thermodynamically distinct from band insulators - there is little to distinguish them in practice.  In terms of their elementary excitations, they are indistinguishable from a simple semiconductor.  A proposal was made in Ref. \cite{yao2007b} for a ``phase-sensitive'' measurement of a d-Mott phase, involving use of a fragile Mott insulator as the insulating component of superconductor-insulator-superconductor junctions of various geometries.  While the experiments involved may not be simple, the analysis highlights another perspective on this new phase.

Let us again consider the checkerboard lattice, initially in the limit $t^\prime \to 0$.  We define a new vacuum state to be the tensor product of the ground-states of each square with two electrons per square.  
We can therefore view \cite{scalapino1996} the ground-state of the extended system as being a Bose insulator with a single d-wave pair per square.  The same Bose insulating state made of s-wave Cooper pairs would likely be adiabatically connected to a band insulator, but in the d-wave case, the insulating state retains information about the internal structure of the pairs.  In the above mentioned experiments, this gives the same sign of any phase-sensitive measurement that would be obtained using a piece of d-wave superconductor as a link in a SQUID loop although of course in the case of the insulator, the Josephson coupling is strongly attenuated.  Needless to say, as with all difficult to realize theoretical proposals these days, a d-Mott phase could be realized for cold atoms in optical lattices under appropriate circumstances \cite{peterson2008,rey2009}.

{\it Acknowledgment}: 
We would like to 
thank Dung-Hai Lee, Srinivas Raghu, Shinsei Ryu, Wei-Feng Tsai, Ashvin Vishwanath, and especially Xiao-Liang Qi 
for insightful discussions. This work was supported, in part, by DOE grants DE-FG02-06ER46287 at Stanford (SAK) and DE-AC02-
05CH11231 at Berkeley (HY).


\begin{thebibliography}{23}
\bibitem{footnote1} We hope to improve this aspect of the proof later.
\end{thebibliography}
\end{document}